\documentclass[epj,nopacs]{svjour}
\usepackage{graphicx} 
\usepackage{amsmath}
\usepackage{bm,amsfonts,amssymb}
\usepackage{color} 
\newcommand{\be}{\begin{equation}}
\newcommand{\ee}{\end{equation}}
\newcommand{\bdm}{\begin{displaymath}}
\newcommand{\edm}{\end{displaymath}}
\newcommand{\upr}{\uparrow}
\newcommand{\dar}{\downarrow}
\newcommand{\barr}{\begin{array}}
\newcommand{\earr}{\end{array}}

\newcommand{\tpz}{${}^{3\!}P_0$}
\newcommand{\tpo}{${}^{3\!}P_1$}

\newcommand{\otpo}{$1\,{}^{3\!}P_1$}
\newcommand{\ttpo}{$2\,{}^{3\!}P_1$}

\newcommand{\tspo}{$2\,{}^{1\!}P_1$}

\newcommand{\otpz}{$1\,{}^{3\!}P_0$}

\newcommand{\osdt}{$1\,{}^{1\!}D_2$}

\newcommand{\jpsi}{J\!/\!\psi}
\DeclareMathAlphabet{\mathitbf}{OML}{cmm}{b}{it}
\begin{document}
\title{$\bm{X(3872)}$ is not a true molecule}
\author{Susana Coito\inst{1} \and George Rupp\inst{1} \and
       Eef van Beveren\inst{2}
}                     
\institute{
Centro de F\'{\i}sica das Interac\c{c}\~{o}es Fundamentais,
Instituto Superior T\'{e}cnico, Technical University of Lisbon,
P-1049-001 Lisbon, Portugal
\and
Centro de F\'{\i}sica Computacional,
Departamento de F\'{\i}sica, Universidade de Coimbra,
P-3004-516 Coimbra, Portugal
}
\date{Received: date / Revised version: date}
%
\abstract{
A solvable coordinate-space model is employed to study the $c\bar{c}$ component
of the $X(3872)$ wave function, by coupling a confined \tpo\ $c\bar{c}$ state
to the almost unbound $S$-wave $D^0\overline{D}^{*0}$ channel via the \tpz\
mechanism. The two-component wave function is calculated for different values
of the binding energy and the transition radius $a$, always resulting in a
significant $c\bar{c}$ component. However, the long tail of the
$D^0\overline{D}^{*0}$ wave function, in the case of small binding,
strongly limits the $c\bar{c}$ probability, which roughly lies in the range
7--11\%, for the average experimental binding energy of 0.16~MeV and $a$
between 2 and 3~GeV$^{-1}$. Furthermore, a reasonable value of 7.8~fm is
obtained for the $X(3872)$ r.m.s.\ radius at the latter binding energy, as
well as an $S$-wave $D^0\overline{D}^{*0}$
scattering length of 11.6~fm. Finally, the $\mathcal{S}$-matrix pole
trajectories as a function of coupling constant show that $X(3872)$ can be
generated either as a dynamical pole or as one connected to the bare $c\bar{c}$
confinement spectrum, depending on details of the model. From these results we
conclude that $X(3872)$ is not a genuine meson-meson molecule, nor actually any
other mesonic system with non-exotic quantum numbers, due to inevitable
mixing with the corresponding quark-antiquark states.
}   


\maketitle

\section{Introduction}
\label{intro}
The Belle Collaboration discovered \cite{PRL91p262001} the charmon\-ium-like
state $X(3872)$ almost a decade ago, observing it in the decay
$B^{\pm} \rightarrow K^{\pm} \pi^+\pi^- J/\psi$, with a significance in excess
of $10\sigma$. The new meson was then confirmed by CDF \cite{PRL93p072001},
D0 \cite{PRL93p162002}, BABAR \cite{PRD71p071103}, and recently also by
LHCb \cite{EPJC72p1972}. In the 2012 PDG tables \cite{PRD86p010001}, $X(3872)$
is listed with a mass of $3871.68\pm0.17$~MeV and a width $\Gamma\!<\!1.2$~MeV.
As for the quantum numbers, the possibilities are $J^{PC}=1^{++}$ and
$J^{PC}=2^{-+}$ according to the PDG \cite{PRD86p010001}. The positive
$C$-parity assignment resulted from an analysis of angular
distributions by CDF \cite{PRL98p132002}. Recently, observation of the decay
to $\gamma J/\psi$ by Belle \cite{PRL107p091803} unambiguously confirmed $C=+$.
For further experiments on $X(3872)$ production, see Ref.~\cite{PRD86p010001}.

On the theory side, the discussion about the nature of $X(3872)$ continues
most vivid. Although the PDG \cite{PRD86p010001} lists $1^{++}$ and $2^{-+}$
as the meson's possible quantum numbers, and a recent BABAR analysis
\cite{PRD82p011101} of the $3\pi$ invariant mass distribution in the decay
$X(3872)\to\omega\jpsi$ even seems to slightly favor the $2^{-+}$ assignment,
most model builders describe the state as an axial vector. For instance, model
calculations of semi-inclusive $B\to\eta_{c2}+X$ processes \cite{PRD85p034032}
as well as electromagnetic $\eta_{c2}$ decays \cite{1205.5725} have been shown
to disfavor the $2^{-+}$ scenario. The same conclusion was reached in a
tetraquark description of $X(3872)$ \cite{CTP57p1033}, while pion exchange in
a molecular picture would be repulsive in this case \cite{PLB590p209} and so
inhibitive of a bound state. Finally, unquenching a \osdt\ $c\bar{c}$ state 
by including meson-meson loops can only further lower the bare mass, which
lies in the range 3.79--3.84~GeV for all quenched quark models we know of,
thus making a $2^{-+}$ charmonium resonance at 3.872~GeV very unlikely
\cite{EPJC71p1762}. For further information and more references concerning
$X(3872)$, see e.g.\ a recent review \cite{PPNP67p390}, as well as our prior
coupled-channel analysis \cite{EPJC71p1762}.

The first suggestion of possible meson-meson molecules bound by pion exchange,
in particular a $DD^*$\footnote
{Henceforth, we omit the bar in $D\overline{D}^*$, for notational
simplicity.}
state with quantum numbers $1^{++}$ or $0^{-+}$, was due to T\"{o}rnqvist
\cite{PLB590p209}.
With the discovery of $X(3872)$ just below the $D^0D^{*0}$ threshold,
this idea was revived, of course. In the present paper, we intend to study
the issue, not from T\"ornqvist's pion-exchange point of view, but rather as
regards its possible implications for models based on quark degrees of
freedom. In this context, it is worthwhile to quote from
Ref.~\cite{PRD76p094028}, in which a molecular interpretation is advocated
(also see Ref.~\cite{PRD69p074005}):
\begin{quote} \em
``Independent of the original mechanism for the resonance,
the strong coupling transforms the resonance into a
bound state just below the two-particle threshold if $a>0$
or into a virtual state just above the two-particle threshold if $a<0$.
If $a>0$, the bound state has a molecular structure,
with the particles having a large mean separation of order $a$.'' \em
\end{quote}
(Note that, here, $a$ is the $S$-wave scattering length.)
Also:
\begin{quote} \em
``In this case \em [$1^{++}$], \em
the measured mass $M_X$ implies unambiguously that $X$ must be either a
charm meson molecule or a virtual state of charm mesons.'' \em
\end{quote}
In face of these peremptory claims about the molecular picture, we believe
it is of utmost importance to study in detail the $X(3872)$ wave function
for a model in which the {\em mechanism} \/generating the meson is quark
confinement combined with strong decay. Thus, we employ a simplified,
coordinate-space version of the coupled-channel model used
recently \cite{EPJC71p1762} to describe $X(3872)$ as a unitarized and
mass-shifted \ttpo\ charmonium state. The model's exact solvability then
allows to obtain analytic expressions for the wave-function components,
and follow bound-state as well as resonance poles on different Riemann
sheets.

The model is outlined in Sec.~\ref{model}, with details moved to
Appendices \ref{appA} and \ref{appB}. Section~\ref{poles} is devoted to
$\mathcal{S}$-matrix pole trajectories as a function of the two free
parameters. In Sec.~\ref{w-f} the two-component wave function is analyzed
for several parameter sets, and is then used in Sec.~\ref{obs} to compute
$c\bar{c}$ probabilities and root-mean-square (r.m.s.) radii. The dynamical
vs.\ confinement nature of the poles is discussed in Sec.~\ref{secdvsc}.
Conclusions are drawn in Sec.~\ref{conc}.

\section{The coupled $\mathitbf{c\bar{c}}\,$-$\mathitbf{D^0\!D^{*0}}$ system}
\label{model}
In our previous work on $X(3872)$ \cite{EPJC71p1762}, we described the state
as a unitarized radially excited $1^{++}$ charmonium resonance, by employing
the resonance-spectrum-expansion (RSE) \cite{AOP324p1620} formalism. The RSE
description of $X(3872)$ amounted to coupling the most relevant meson-meson
channels allowed by the Okubo-Zweig-Izuka (OZI) rule to a spectrum of bare
$1^{++}$ $c\bar{c}$ states, and also the OZI-suppressed but experimentally
observed \cite{PRD86p010001} $\rho^0\jpsi$ and $\omega\jpsi$ channels. Thus,
we found that unquenching shifts the bare \ttpo\ state more than 100 MeV
downwards, which settles as a very narrow resonance slightly below or on top
of the $D^0D^{*0}$ threshold. However, the RSE approach does not allow to
obtain wave functions in a straightforward fashion. So for the purpose of the
present paper, we resort to the equivalent \cite{IJTPGTNLO11p179}
coordinate-space coupled-channel formalism of Ref.~\cite{ZPC19p275}, which was
used to study the influence of strong decay channels on hadronic spectra
and wave functions, besides several more specific phenomenological
applications. Furthermore, since here we do not aim to describe all aspects of
the $X(3872)$ resonance in a detailed way, but rather want to focus on the
importance of the charmonium component in the wave function, we restrict
ourselves henceforth to a simple two-channel system, viz.\ a \tpo\ $c\bar{c}$
state coupled to an $S$-wave $D^0D^{*0}$ channel, the latter being the dominant
one, observed in the $D^0\overline{D}^0\pi^0$ mode \cite{PRD86p010001}.
A partial study of this problem was already carried out in Ref.~\cite{1209.1313}.

Consider now a system composed of a confined $q\bar{q}$ channel coupled to a
meson-meson channel $M_1M_2$. Confinement is described through a
harmonic-oscillator (HO) potential with constant frequency $\omega$,
having spectrum
\be
\label{ho}
E=(2\nu+l_c+\frac{3}{2})\,\omega+m_q+m_{\bar{q}} \; ,
\ee
where $\nu$ is the radial quantum number, $l_c$ the $q\bar{q}$ orbital
angular momentum, and $m_q=m_{\bar{q}}=2\mu_q$ the constituent quark mass.
In the scattering channel, no direct interactions between
the two mesons are considered, with $\mu_f$
and $l_f$ being the reduced two-meson mass and orbital angular momenta
in the free channel, respectively.
Transitions between the two channels are modeled via an off-diagonal
delta-shell potential with strength $g$, which mimics string breaking at a
well-defined distance $a$. The corresponding Hamiltonian, transition
potential, and $2\times2$ matrix Schr\"odinger equation are given in
Eqs.~(\ref{hc})--(\ref{schr}), with the usual definition $u(r)=rR(r)$, where
$R(r)$ is the radial wave function. The exact solution to these equations
is derived in Appendix \ref{appA}:
\be
\label{hc}
h_c=\frac{1}{2\mu_c}\bigg(-\frac{d^ 2}{dr^ 2}+\frac{l_c(l_c+1)}
{r^2}\bigg)+\frac{\mu_c\omega^2r^2}{2}+m_q+m_{\bar{q}} \; ,
\ee
\be
\label{hf}
h_f=\frac{1}{2\mu_f}\bigg(-\frac{d^ 2}
{dr^ 2}+\frac{l_f(l_f+1)}{r^2}\bigg)+m_{M_1}+m_{M_2} \; ,
\ee
\be
\label{pot}
V=\frac{g}{2\mu_ca}\delta(r-a) \; ,
\ee
\be
\label{schr}
\left(\barr{cc}
h_c & V\\
V & h_f
\earr\right)
\left(\barr{c}
u_c\\
u_f
\earr\right)=
E\left(\barr{c}
u_c\\
u_f
\earr\right) \; .
\ee

Once the $1\times1$ $\mathcal{S}$ matrix (cf.\ Eq.~(\ref{scotan})) has been
constructed from the wave function, possible bound or virtual states as well
as resonances can be searched for. These occur at real or complex energies
for which $\mathcal{S}$ blows up, or equivalently, when
$\cot \delta_{l_f}(E)=i$ (cf.\ Eq.~(\ref{cotan})). Thus, unlike the pure HO
spectrum, which is a real and discrete set of energies, the ``unquenched''
spectrum given by Eq.~(\ref{schr}) includes complex energies, too, some of
which even have no obvious connection to the original, ``bare'' levels. The 
corresponding poles, called {\em dynamical}, are the result of attraction 
in the meson-meson channel generated by the interaction with intermediate
quark-antiquark states. The light scalar mesons $f_0(500)$, $K_0^*(800)$,
$f_0(980)$, and $a_0(980)$ \cite{PRD86p010001} are archetypes of such
resonances \cite{ZPC30p615}. On the other hand, we designate by
{\it confinement poles} \/ the ones that can be linked straightforwardly to
the bare states. Nevertheless, we shall show below --- as we already
demonstrated in previous work --- that these two types of resonances are not
so distinct after all, since minor parameter changes may transform one kind
into the other. But regardless of the type of pole, the corresponding radial
quantum number in Eq.~(\ref{eve}) is related to the energy of the 
coupled-channel system by
\be
\label{eve}
\nu(E)=\frac{E-2m_c}{2\omega}-\frac{l_c+3/2}{2} .
\ee 
Only in the uncoupled case, that is, for $g=0$, one recovers the original
HO spectrum, with $\nu=0,1,2,...$. 

Now we apply the formalism to the coupled $c\bar{c}$-$D^0D^{*0}$ system.
The $c\bar{c}$ channel is assumed to be in a \tpo\ state, i.e., with $l_c=1$,
implying the $D^0D^{*0}$ channel to have $l_f=0$ or 2. Nevertheless, we shall
restrict ourselves here to the $S$-wave channel only, which will be strongly
dominant, especially near threshold. The fixed parameters are given in
Table~\ref{tab:param}, where the meson masses are from the
\begin{table}[!b]
\centering
\caption{\label{tab:param} Fixed parameters.}
\begin{tabular}{c|c|c|c|c|c}
Param.\ & $\omega$&$m_c$&$m_{D^0}$&$m_{D^{*0}}$&$m_{D^0}\!+\!m_{D^{*0}}$\\
\hline \mbox{ }\\[-9pt]
(MeV)&$190$&$1562$&$1864.86$&$2006.98$&$\bf{3871.84}$\\
\end{tabular}
\end{table}
PDG \cite{PRD86p010001}, while the HO frequency $\omega$ and the constituent
charm quark mass $m_c$ are as originally determined in
Ref.~\cite{PRD27p1527} and left unaltered ever since. Thus,
from Eq.~(\ref{ho}) we get the lowest two HO states at $E_0=3599$~MeV and
$E_1=3979$ MeV, respectively. The former should give rise --- after
unquenching --- to the \otpo\ charmonium state $\chi_{c1}(1P)$
\cite{PRD86p010001}, with mass $3511$ MeV, while the latter is the bare
\ttpo\ state, which cannot so easily be linked to resonances
in the PDG tables, though both $X(3940)$ and $X(3872)$ are possible
candidates, in view of their mass and dominant $DD^*$ decay mode
\cite{PRD86p010001}. However, $X(3940)$ may just as well be the, so far
unconfirmed, \tspo\ ($1^{+-}$) state $h_c(2P)$ \cite{EPJC71p1762}.

The two remaining parameters, viz.\ the string-breaking distance $a$
and the global coupling $g$, have to be adjusted to the experimental data.
Nevertheless, these parameters are not completely free, as they both have a
clear physical interpretation, albeit of an empirical nature. Thus, $a$ is
the average interquark separation at which \tpo\ quark-pair
creation/annihilation is supposed to take place, while $g$ is the overall
coupling strength for such processes. Note that we do not assume a particular
microscopic model for string breaking inspired by QCD, like e.g.\ in a
very recent paper \cite{1210.4674}. Still, the values of $a$ found in
the present work are in rough agreement with our prior model findings, and
even compatible \cite{0712.1771} with a lattice study of string breaking in
QCD \cite{PRD71p114513}. Concretely, we have been obtaining values of $a$
in the range 1--4 GeV$^{-1}$ (0.2--0.8~fm), logically dependent on quark
flavor, since the string-breaking distance will scale with the meson's size,
being smallest for bottomonium.
As for the coupling parameter $g$, its empirical value will depend on $a$,
but also on the set of included decay channels. In
realistic calculations, values of the order of 3 have been obtained (see
e.g.\ our previous paper \cite{EPJC71p1762} on $X(3872)$).

\section{Poles}
\label{poles}
The crucial test the present model must pass is its capability of generating
a pole near the $D^0D^{*0}$ threshold. Indeed, a dynamical pole is
found slightly below threshold for different combinations of the free
parameters $a$ and $g$, several of which are listed in Table~\ref{ag}.
\begin{table}[!b]
\caption{\label{ag} Bound states (BS), virtual bound states (VBS), and
resonances closest to threshold, for various $g$ and $a$ combinations.}
\begin{center}
\begin{tabular}{c|c|l|c}
$a$ (GeV$^{-1}$)& $g$  & pole (MeV) & type\\
\hline \mbox{ } \\[-9pt]
$2.0$ & $1.149$ & $3871.84$&VBS\\
$2.5$ & $1.371$ & $3871.84$&VBS\\
$3.0$ & $2.142$ & $3871.84$&VBS\\
$3.1$ & $2.503$ & $3871.84$&VBS\\
$3.2$ & $2.531$ & $3871.84-i12.01$&resonance\\
$3.3$ & $3.723$ & $3871.84-i\ 4.45$&resonance\\
$3.4$ & $7.975$ & $3871.84-i\ 0.39$&resonance\\
$3.5$ & $\infty$ &  & - \\
\hline \mbox{ } \\[-9pt]
$2.0$ & $1.152$ & $3871.84$&BS\\
$2.5$ & $1.373$ & $3871.84$&BS\\
$3.0$ & $2.145$ & $3871.84$&BS\\
$3.1$ & $2.507$ & $3871.84$&BS\\
$3.2$ & $3.083$ & $3871.84$&BS\\
$3.3$ & $4.194$ & $3871.84$&BS\\
$3.4$ & $8.254$ & $3871.84$&BS\\
$3.5$ & $\infty$ &  & - \\
\end{tabular}
\end{center}
\end{table}
Examples are here given of bound states, virtual bound states,
and below-threshold resonances, the latter ones only occurring for $S$-wave
thresholds as in our case. Note that poles of both virtual bound states and
resonances lie on the second Riemann sheet, i.e., the relative momentum has a
negative imaginary part. From this table we also observe that larger and larger
couplings are needed to generate a pole close to threshold when $a$ approaches
the value 3.5~GeV$^{-1}$. We shall see below that this is due to the nodal
structure of the bound-state wave function.

Although a dynamical pole shows up near the $D^0D^{*0}$ threshold, there still
should be a confinement pole connected to the first radial \tpo\ excitation at
3979~MeV. Well, we do find such a pole, for each entry in Table~\ref{ag}. In
Table~\ref{twopoles} a few cases are collected, with the parameters tuned to 
\begin{table}[!t]
\centering
\caption{\label{twopoles} Pole doubling: pairs of poles (in MeV) for some sets
of $a$ and $g$ values, chosen such that the dynamical pole settles at
the $X(3872)$ PDG \cite{PRD86p010001} mass.}
\begin{tabular}{c|c|c|c}
$a$ (GeV$^{-1}$) & $g$ & dynamical pole & confinement pole\\
\hline \mbox{ } \\[-9pt]
2.0 & 1.172 & 3871.68 & $4030.50 - i 136.51$\\
2.5 & 1.403 & 3871.68 & $4063.27 - i 124.07$\\
3.0 & 2.204 & 3871.68 & $4101.48 - i\ 88.03$\\
3.4 & 8.623 & 3871.68 & $4185.85 - i\ 20.63$\\
\end{tabular}
\end{table}
generate a dynamical pole at precisely the $X(3872)$ PDG \cite{PRD86p010001}
mass of 3871.68~MeV. Note, however, that the associated confinement pole is not
necessarily of physical relevance, since at the corresponding energy several
other strong decay channels are open, which no doubt will have a very
considerable influence and possibly even change the nature of both poles. As
a matter of fact, in our prior paper \cite{EPJC71p1762}, with all relevant
two-meson channels included, the $X(3872)$ resonance was found as a confinement
pole, whereas dynamical poles were only encountered very deep in the complex
energy plane, without any observable effect at real energies. So here we show
these results only to illustrate that pole doubling may occur when strongly
coupling $S$-wave thresholds are involved, as we have observed in the past 
in the case of e.g.\ the light scalar mesons \cite{ZPC30p615} and 
$D_{s0}^*(2317)$ \cite{PRL91p012003}. The issue of confinement vs.\ dynamical
poles will be further studied in Sec.~\ref{secdvsc}. 

\begin{figure}[!t]
\centering
\caption{\label{traja2}Pole trajectories of dynamical (left) and
confinement (right) poles as a function of $g$, for $a$=2.0 GeV$^{-1}$ (top),
3.0 GeV$^{-1}$ (middle), and 3.5 GeV$^{-1}$ (bottom), respectively.
In the last case, there is no bound state near threshold. Note: (i)
poles in Table.~\ref{twopoles} are here marked by
\boldmath$\ast$; (ii) arrows along curves indicate increasing $g$.}
\begin{tabular}{c c }
\hspace*{-20pt}
\resizebox{!}{400pt}{\includegraphics{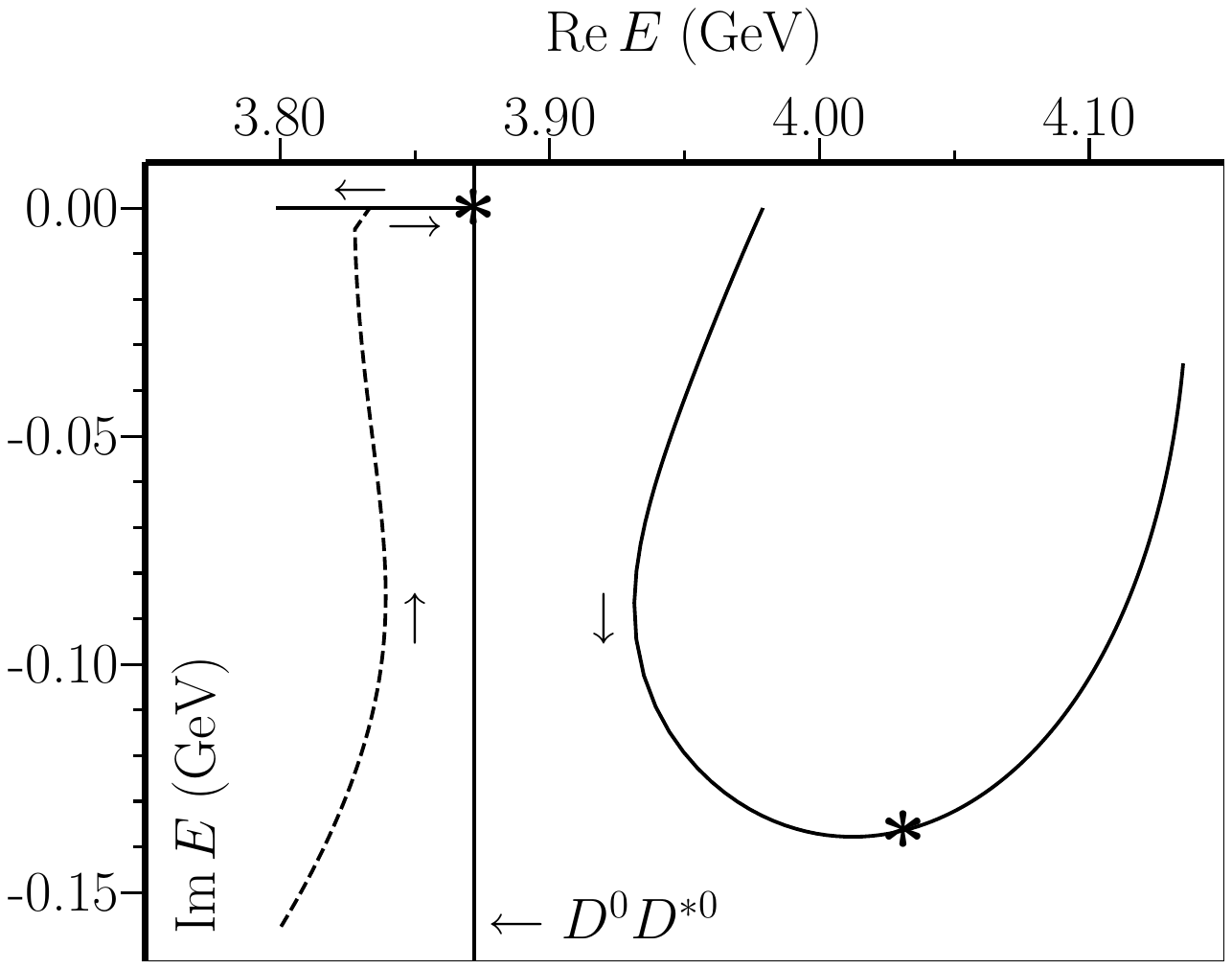}}\\
\mbox{} \\[-90mm]
\hspace*{-20pt}\resizebox{!}{400pt}{\includegraphics{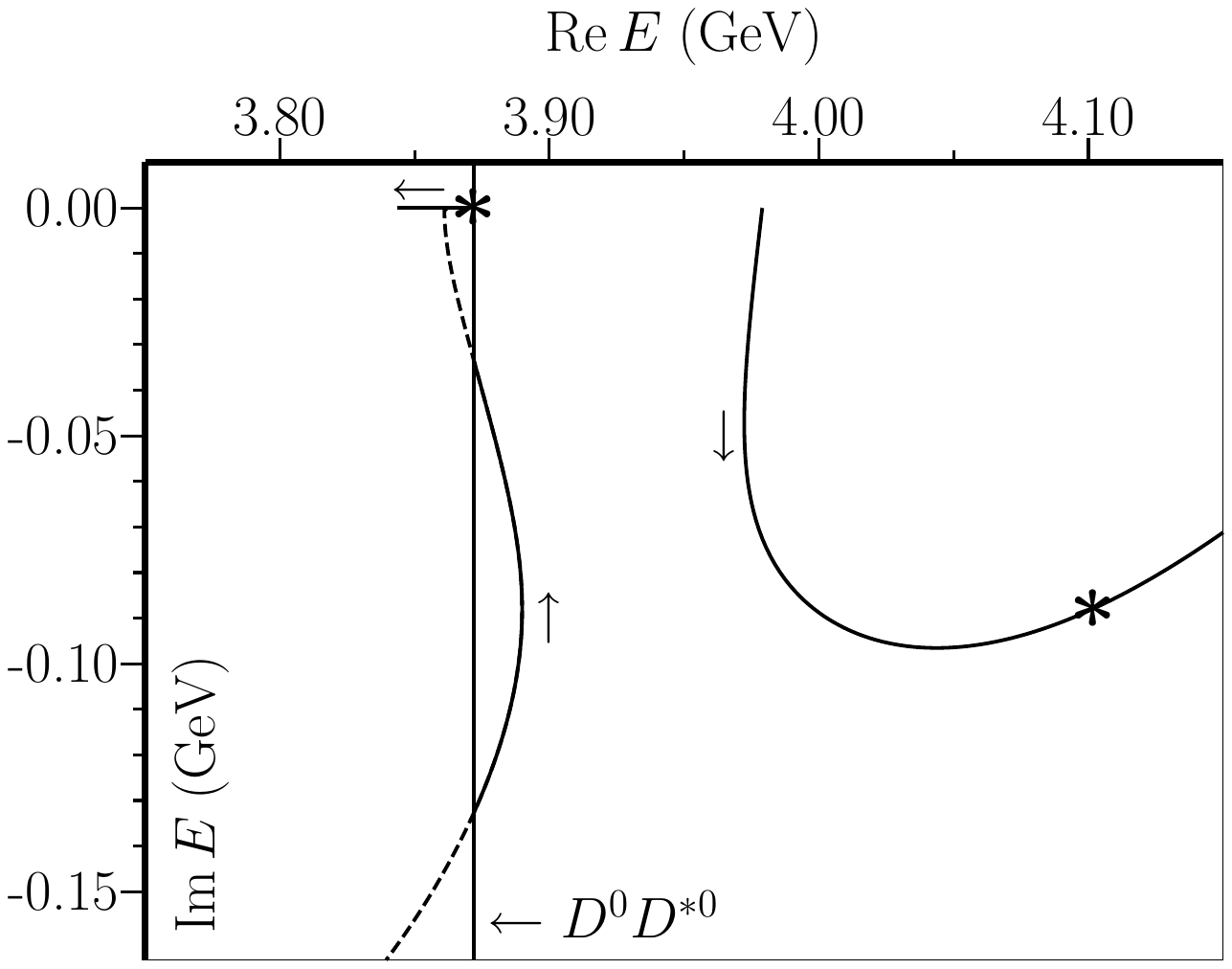}}\\
\mbox{} \\[-90mm]
\hspace*{-20pt}\resizebox{!}{400pt}{\includegraphics{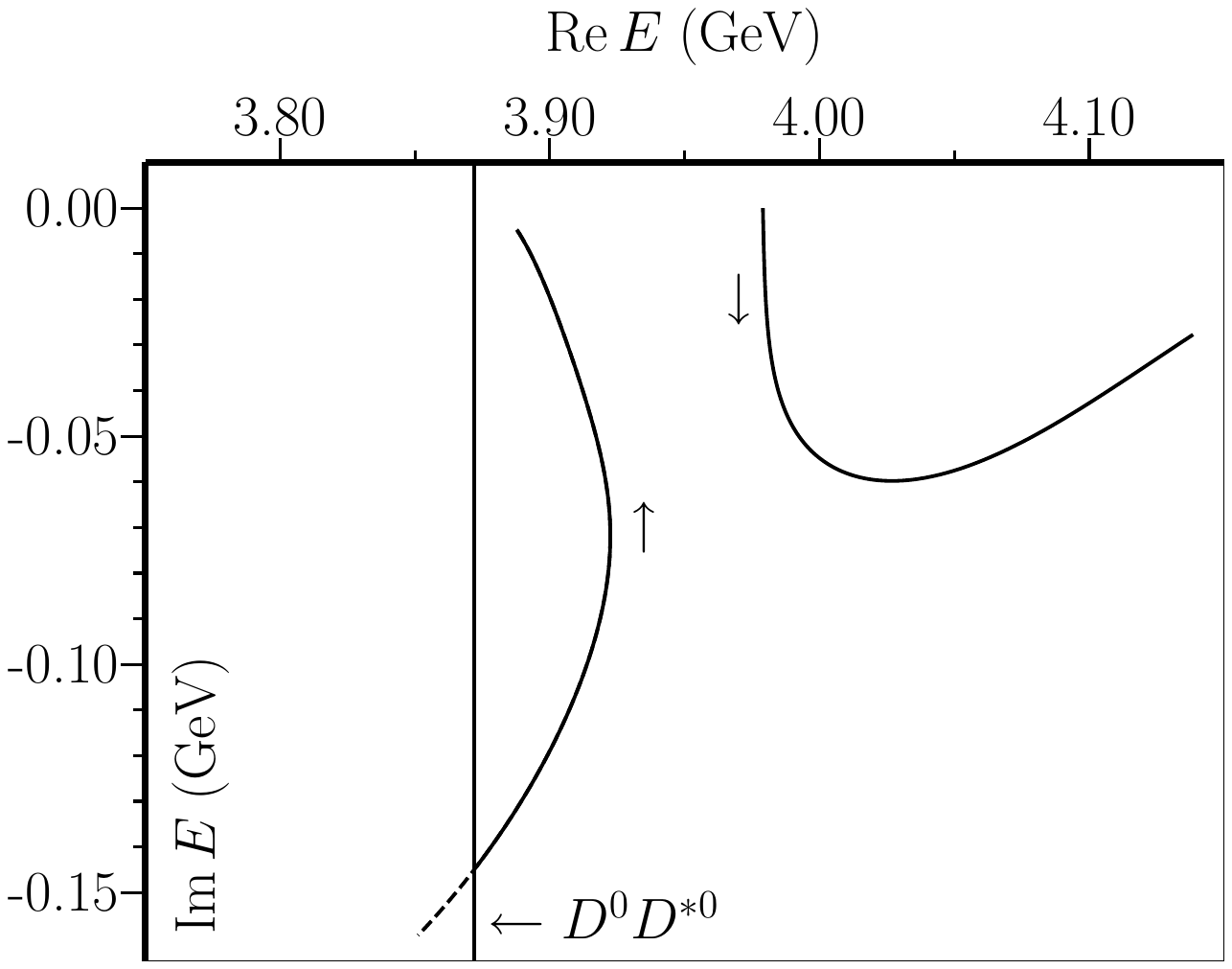}}\\
\mbox{} \\[-90mm]
\end{tabular}
\end{figure}

In order to better understand the dynamics of the different poles, we plot
in Fig.~\ref{traja2} pole trajectories in the complex energy plane as a function
of the coupling constant $g$, and for three different values of $a$. For
vanishing $g$, the dynamical pole acquires a negative infinite imaginary part
and so disappears in the continuum, whereas the confinement pole moves to the 
real energy level of the bare \ttpo\ state, i.e., 3979~MeV. As $g$ increases,
and for both $a=2.0$ GeV$^{-1}$ and $a=3.0$ GeV$^{-1}$, the dynamical pole 
moves to the real axis below threshold, becoming first a virtual bound
state and then a genuine bound state. Note that, in the latter case, the
real part twice attains the $X(3872)$ mass even before the pole reaches
the real axis, but the corresponding imaginary parts are much too
large as compared with experiment \cite{PRD86p010001}, so only the bound
state can be considered physical. Finally, for $a=3.5$~GeV$^{-1}$ the pole
does never reach the real axis, which would require an infinite coupling.
For the other parameter sets listed in Table~\ref{ag}, we find intermediate
situations. Another feature we can observe for all trajectories is an initial
attraction and subsequent repulsion between the dynamical and the confinement
poles.

\section{Wave function}
\label{w-f}
\begin{figure}[hb]
\centering
\caption{\label{fig:rwf}Normalized two-component radial wave function $R(r)$
for three BEs, corresponding to labels $A,X,D$ in
Table~\ref{binden}, and two $a$ values. Upper curves: $R_f(r)$; lower curves:
$R_c(r)$. Left: $a=2$~GeV$^{-1}$; right: $a=3$~GeV$^{-1}$.}
\begin{tabular}{c c}
\hspace*{-95pt}
\resizebox{!}{430pt}{\includegraphics{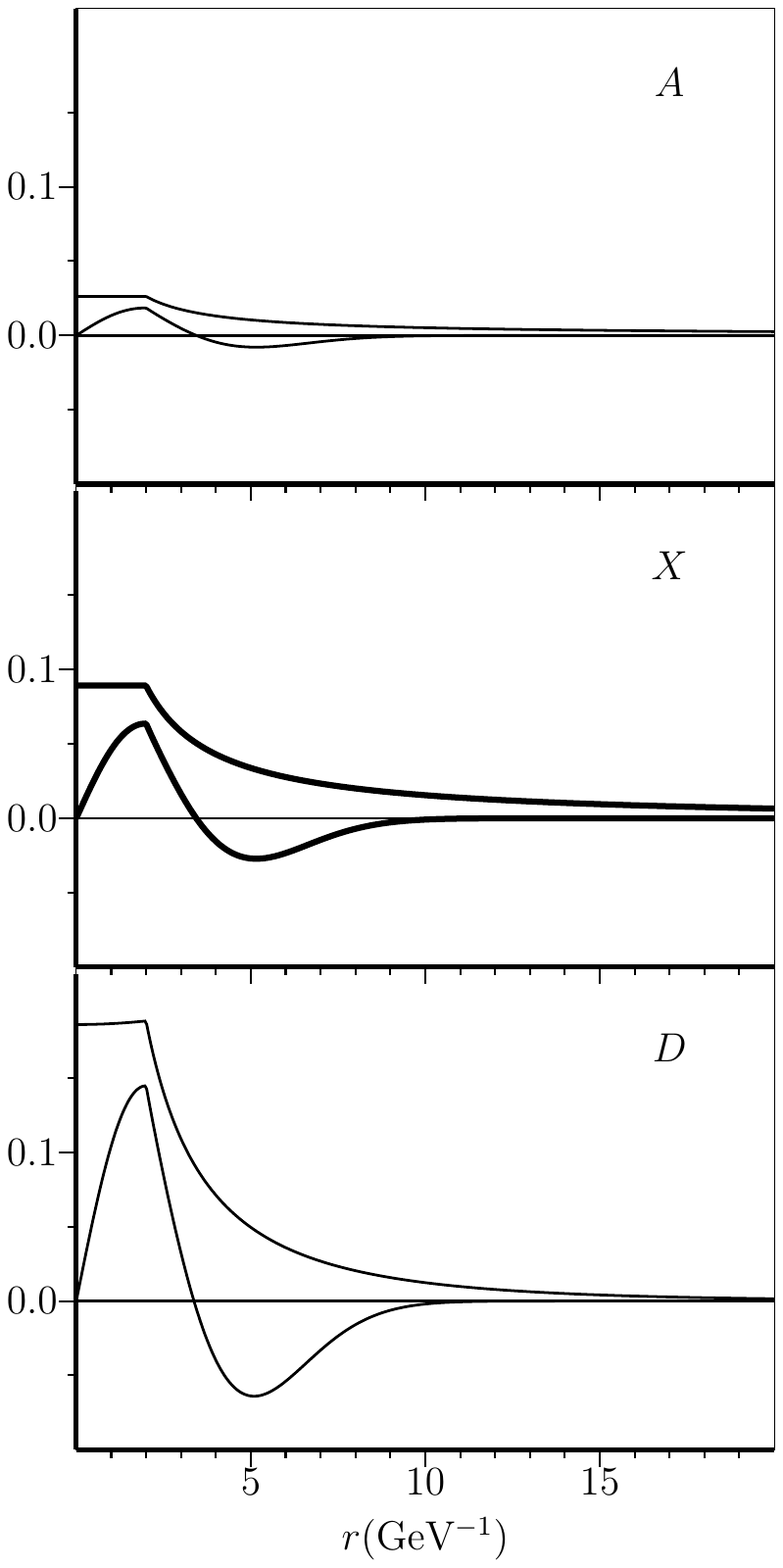}}&
\hspace*{-215pt}\resizebox{!}{430pt}{\includegraphics{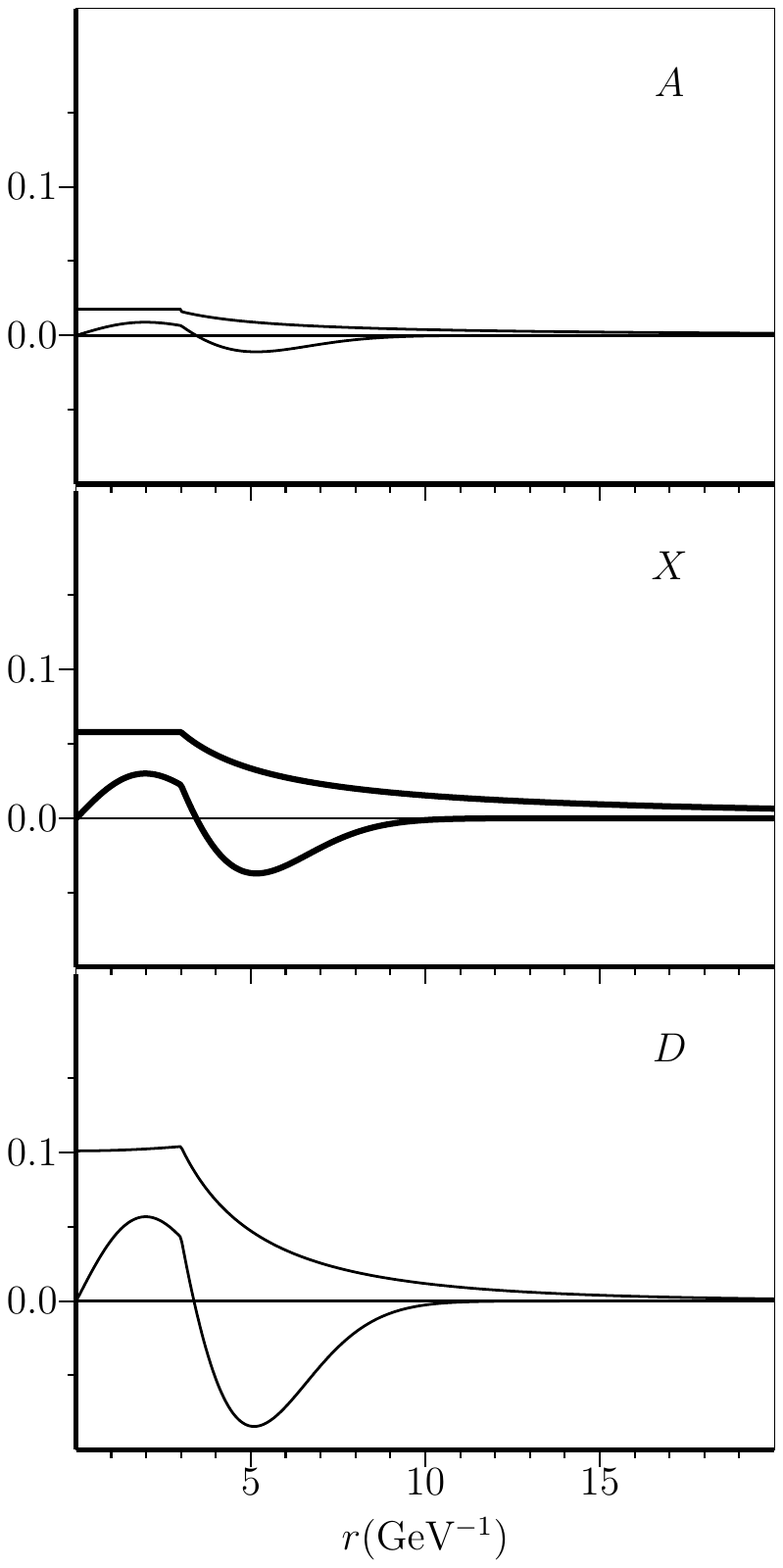}}\\
\end{tabular}
\mbox{} \\[-70mm]
\end{figure}

Now we are in a position to study the $X(3872)$ bound-state wave function in
several situations. We choose two values for the string-breaking parameter, 
viz.\ $a=2.0$~GeV$^{-1}$ and $a=3.0$~GeV$^{-1}$. In Table \ref{binden} five
\begin{table}
\centering
\caption{\label{binden}Five chosen binding energies (BE) in the $D^0D^{*0}$
channel, for two different $a$ values and the corresponding couplings $g$.}
\begin{tabular}{c| c| c| c| c|c}
\multicolumn{2}{c|}{$a$ (GeV$^{-1}$)}   & \multicolumn{2}{c|}{2.0} &
\multicolumn{2}{c}{3.0}\\
\hline \mbox{ } \\[-9pt] 
label& BE (MeV) & $g$     & pole    & $g$     & pole \\
\hline \\[-9pt] 
$A$      & $\ 0.00$ & 1.152 & 3871.84 & 2.145 & 3871.84\\
$B$      & $\ 0.10$ & 1.167 & 3871.74 & 2.191 & 3871.74\\
$\mathbf{X}$ & $\mathbf{\ 0.16}$ & \bf{1.172} & \bf{3871.68} & \bf{2.204} & \bf{3871.68}\\
$C$      & $\ 1.00$ & 1.207 & 3870.84 & 2.311 & 3870.84\\
$D$      & $ 10.00$ & 1.373 & 3861.84 & 2.899 & 3861.84
\end{tabular}
\end{table}
different binding energies (BEs) are chosen with respect to the $D^0D^{*0}$
channel, including the PDG \cite{PRD86p010001} value labeled by $X$. We have
computed and normalized (see Appendix~\ref{appA}) the two-component radial
wave function $R(r)$ for each  of the five cases. In
Fig.~\ref{fig:rwf} we depict the cases labeled by $A$, $X$ and $D$, the other
two representing intermediate situations. General features we immediately
observe are the typical $S$-wave behavior of the $D^0D^{*0}$ wave-function
component $R_f$, while the $c\bar{c}$ wave function $R_c$ is in a $P$ state,
the latter also having a node, as it is dominantly a first radial excitation.
Furthermore, $|R_f|$ is larger than $|R_c|$ in most situations, for
all $r$, except for unphysically large BEs (cf.\ plot $D$).
Nevertheless, the two components are of comparable size for intermediate $r$
values. Then, as the BE becomes smaller, the tail of $R_f$ grows
longer, as expected, whereas $R_c$ always becomes negligible for
distances larger than roughly 11--12 GeV$^{-1}$. Now, the increased $R_f$ tail
affects the normalization of \em both \em \/$R_c$ and $R_f$. Thus, the ratio
$|R_f(r)|/|R_c(r)|$ is quite robust for most $r$ values, as it does not
significantly change with the BE.

\section{Probabilities and r.m.s.\ radii}
\label{obs}
Having derived the $X(3872)$ wave function for several scenarios, we can
now straightforwardly compute the relative probabilities of the $c\bar{c}$
and $D^0D^{*0}$ components (see Appendix~\ref{appA}), with the results given
in Table~\ref{prob}, for the
\begin{table}
\centering
\caption{\label{prob} Probabilities (in \%) of the two wave-function
components, for the cases specified in Table~\ref{binden} ($a$ in GeV$^{-1}$).}
\begin{tabular}{ c|c|c|c|c|c|c}
$a$  & channel        & $A$ & $B$ & $\mathbf{X}$ & $C$ & $D$\\
\hline \mbox{ } \\[-9pt]
2.0 &$c\bar{c}$   & $\ 0.63$  & $\ 6.00$ & ${\bf \ 7.48}$ & $16.98$ & $39.68$\\
2.0 &$D^0D^{*0}$  & $ 99.37$  & $ 94.00$ & $ 92.52$       & $83.02$ & $60.32$\\
\hline \mbox{ } \\[-9pt]
3.0 &$c\bar{c}$   & $\ 0.97$  & $\ 9.01$ & ${\bf 11.18}$  & $24.65$ & $55.54$\\
3.0 &$D^0D^{*0}$  & $ 99.03$  & $ 90.99$ & $88.82$        & $75.35$ & $44.46$
\end{tabular}
\end{table}
five BEs and two $a$ values from Table~\ref{binden}. Note that the
probability in the $D^0D^{*0}$ channel is only computed for normalization
purposes, since in a more realistic calculation at least the $D^\pm D^{*\mp}$
component would acquire a nonnegligible probability as well, as the
corresponding threshold lies only 8~MeV higher. Nevertheless, our
simplification is unlikely to have an appreciable effect on the $c\bar{c}$
probability and will only increase that of the $D^0D^{*0}$ component
accordingly. Also note that the $c\bar{c}$ probability includes all \tpo\
states, with the \ttpo\ being dominant, because the corresponding bare
eigenstate lies only 100~MeV higher.  However, also the \otpo\ state is
nonnegligible in the physical $X(3872)$ wave function. In the coupled-channel
approach of Ref.~\cite{PRD85p114002}, a \otpo\ admixture of about 15\% was
found.  Notice that --- inevitably --- unquenching not only mixes meson-meson
components into the total bound-state wave function, but also quark-antiquark
components of confinement states other than the one under consideration (also
see Ref.~\cite{ZPC19p275}). Here, for a BE of 0.16~MeV, corresponding to the
physical \cite{PRD86p010001} $X(3872)$, case $\mathbf{X}$ in Table~\ref{prob},
has a 7.48\% $c\bar{c}$ probability for $a=2.0$ GeV$^{-1}$ and 11.18\% for
$a=3.0$ GeV$^{-1}$. For smaller BEs, the $c\bar{c}$ probability decreases as
expected, because of the growing weight of the $D^0D^{*0}$ tail. On the other
hand, for a BE of 10~MeV and $a=3.0$~GeV$^{-1}$, the charmonium probability
becomes even larger than that of the meson-meson component. Now,
the experimental errors in the average mass of the $X(3872)$ and the
$D^0D^{*0}$ threshold allow for a maximum BE of 0.57~MeV, i.e., somewhere
in between cases $\mathbf{X}$ and $C$. This would then correspond to a
$c\bar{c}$ probability roughly midway in the range 7.48\%--16.98\%
($a=2.0$~GeV$^{-1}$) or 11.18\%--24.65\% ($a=3.0$~GeV$^{-1}$).
In the limiting case of zero binding, the $c\bar{c}$ probability
would eventually vanish. Also notice that, in all five cases of
Table~\ref{prob}, the $c\bar{c}$ probability rises by about 50\%
when $a$ is increased from 2.0 to 3.0~GeV$^{-1}$.  Nevertheless,
if we take $a=2.0$~GeV$^{-1}$ as in our Ref.~\cite{EPJC71p1762},
we get a $c\bar{c}$ probability of 7.48\%, very close the 7\% found
in Refs.~\cite{PRD81p054023,1206.4877}.

Next we use the normalized wave functions and Eq.~(\ref{rms}) to compute
the $X(3872)$ r.m.s.\ radius for the five cases discussed before
(see Table~\ref{binden}), with the results presented in Table~\ref{tab:avr}. 
\begin{table}
\centering
\caption{\label{tab:avr} R.m.s.\ radii of the wave function, expressed in fm,
for the cases specified in Table~\ref{binden}.}
\begin{tabular}{c c| c c c c c}
$a$ (GeV$^{-1}$) & $a$ (fm) & $A$ & $B$ & $\mathbf{X}$ & $C$ & $D$\\
\hline \mbox{ } \\[-9pt]
2.0 & 0.39 & 100.22 & 9.92 & \bf{7.82} & 3.10 & 1.15\\
\hline \mbox{ } \\[-9pt]
3.0 & 0.59 & 100.14 & 9.85 & \bf{7.76} & 3.05 & 1.23 
\end{tabular}
\end{table}
It is interesting to observe that the r.m.s.\ radius, which in principle
is an observable, is much less sensitive to the choice of $a$ than de
wave-function probabilities. Furthermore, the large to very large
r.m.s.\ radii in the various situations are hardly surprising, in view
of the small binding energies and the resulting very long tails of the
$D^0D^{*0}$ wave-function components (see Fig.~\ref{fig:rwf} above).

Using Eq.~(\ref{cotan}), we now also evaluate the $S$-wave scattering length
$a_S=-\lim_{E\to0}\left[k(E)\cot\delta_0(E)\right]^{-1}$. In case
$\mathbf{X}$ and for $a=2.0$ GeV$^{-1}$ we thus find $a_S=11.55$ fm, which is
large yet of the expected order of magnitude for a BE of 0.16 MeV. For even
smaller BEs, the scattering length will further increase, roughly like
$\propto\!1/\sqrt{\mbox{BE}}$. Let us here quote from Ref.~\cite{PRD69p074005}:
\begin{quote} \em
``Low-energy universality implies that as the scattering length $a$ increases,
the probabilities for states other than $D^0\bar{D}^{*0}$ or
$\bar{D}^0D^{*0}$ decrease as $1/a$ \ldots''
\end{quote}
Indeed, we verify from our Table~\ref{prob} that  --- very roughly ---
the $c\bar{c}$ probability decreases as $\propto\!\sqrt{\mbox{BE}}$,
and so like $\propto\!1/a_S$.

\section{Stability of results and nature of poles}
\label{secdvsc}
\begin{figure}[h!]
\centering
\caption{\label{newtraj} Trajectories of dynamical and confinement poles.
The bold curves represent cases $I$ \/(top graph) and $II$ \/(bottom graph)
defined in Table~\ref{tab:newpar}, and the others the standard case of
Fig.~\ref{traja2}; the solid (dashed) lines stand for normal (below-threshold)
resonances. All trajectories lie on the second Riemann sheet. The pole
positions for the $g$ values in Table~\ref{tab:newpar} are marked by
$\mathbf{\ast}$.}
\begin{tabular}{c c}
\hspace*{-40pt}
\resizebox{!}{450pt}{\includegraphics{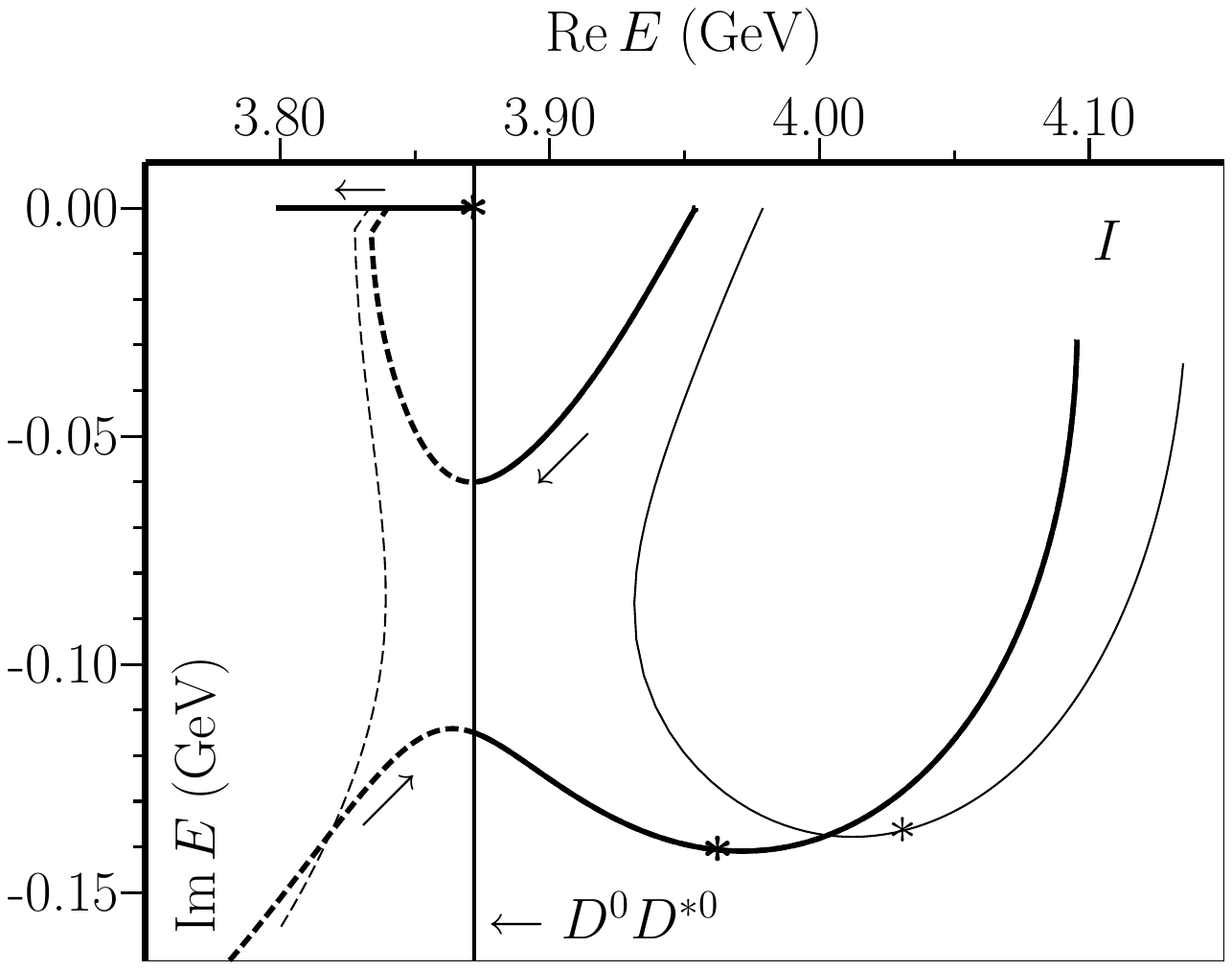}}\\
\mbox{} \\[-95mm]
\hspace*{-40pt}\resizebox{!}{450pt}{\includegraphics{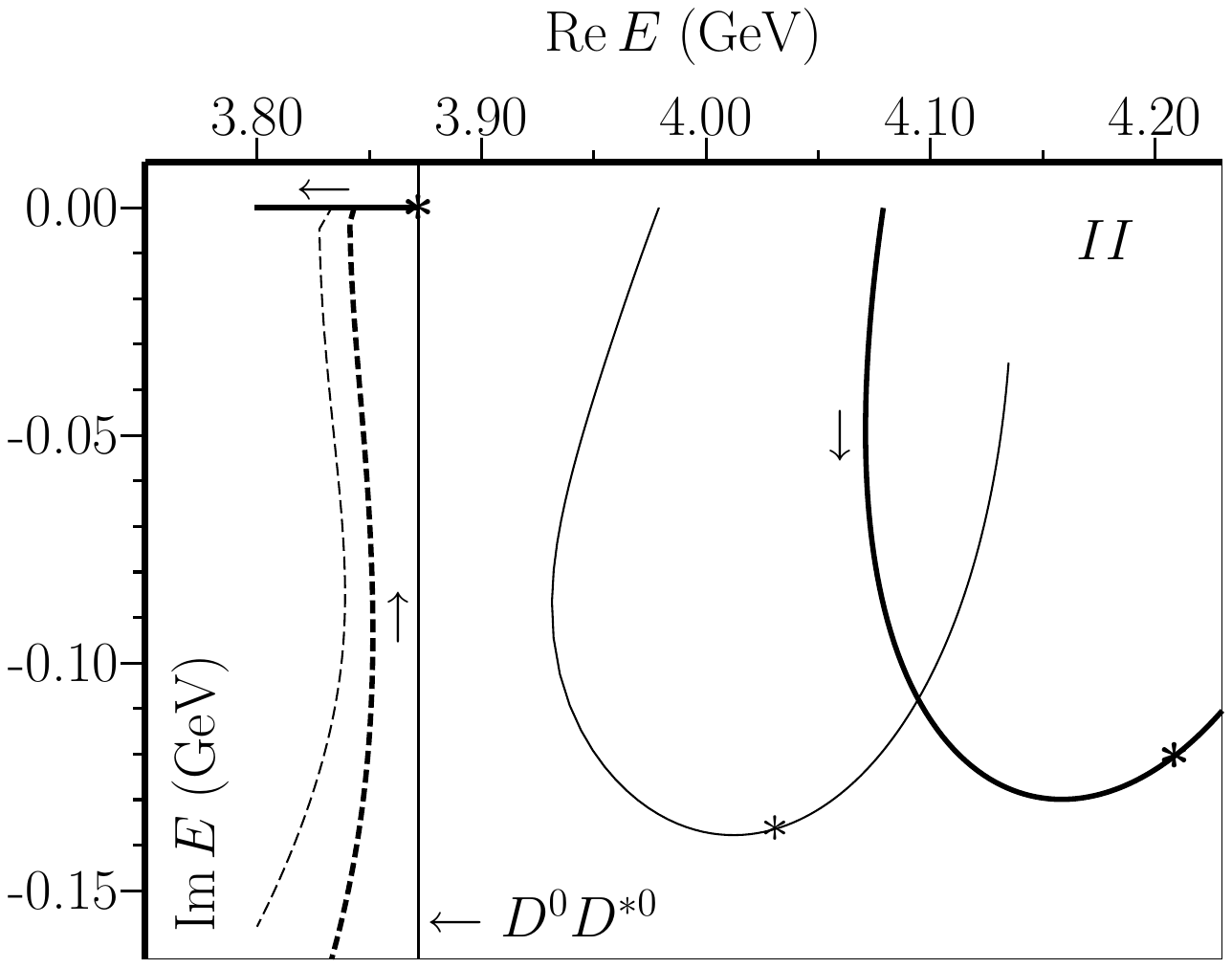}}\\
\mbox{} \\[-100mm]
\end{tabular}
\end{figure}

In this section we are going to study the stability of our results, as well as
the nature of the found solutions. So let us
vary the two usually fixed parameters, viz.\ $\omega$ and $m_c$, in such a
way that the bare \otpo\ mass remains unaltered at 3599~MeV, whereas that of the
\ttpo\ changes as shown in Table.~\ref{tab:newpar}.
\begin{table}[b!]
\centering
\caption{\label{tab:newpar}Probability of $c\bar{c}$ component and $X(3872)$
r.m.s.\ radius for varying $\omega,m_c$, with bare $E_0$ fixed at 3599~MeV,
$X(3872)$ pole at 3871.68~MeV, and $a=2.0$~GeV$^{-1}$.}
\begin{tabular}{c|c c c}
            &  $I$ &  standard   &  $II$\\
\hline \mbox{ } \\[-9pt]
$E_1$ (MeV) & 3954 & 3979 & 4079\\
\hline \mbox{ } \\[-9pt]
$m_c$ (MeV)       & 1577.63 & 1562  & 1499.5\\
$\omega$ (MeV)    & 177.5   & 190   & 240\\
$g$               & 1.034   & 1.172 & 1.572\\
$c\bar{c}$ ($\%$) & 9.49 & $\mathbf{7.48}$ & 6.51 \\
$r_{\mbox{\scriptsize r.m.s.}}$ (fm)    & 7.72 & 7.82 & 8.83\\
\end{tabular}
\end{table}
Thus, in case $I$ $E_1$ is lowered by $25$~MeV, while in case $II$ it rises
by $100$ MeV. The trajectories for these two new situations are plotted in
Fig.~\ref{newtraj}.
For $I$ \/we observe that, just as in the standard case depicted
in Fig.~\ref{traja2}, two poles are found relatively close to the
real axis, of a dynamical and a confinement origin, respectively. However,
now it is the \ttpo\ confinement pole that moves steadily downwards
and settles on the real axis below threshold, whereas the dynamical
pole moves to higher energies and eventually approaches the real axis. So
the poles interchange their roles when going from the standard case to
case $I$. Nevertheless, the values of $g$ needed to get a bound state
at 3871.68~MeV are not very different in the two cases, viz.\ 1.172 vs.\ 1.034.
Such a behavior was already observed almost a decade ago, namely for
$D_{s0}^*(2317)$ \cite{PRD86p010001} charmed-strange meson. In a first,
two-channel model calculation \cite{PRL91p012003} the $D_{s0}^*(2317)$ showed
up as a dynamical resonance, settling below the $S$-wave $DK$ threshold, 
whereas the \otpz\ $c\bar{s}$ state turned out to move to higher energies,
with a large width, similarly to the standard $X(3872)$ case in
Figs.~\ref{traja2} and \ref{newtraj} above. However, in a more complete,
multichannel approach \cite{PRL97p202001} the situations got reversed,
just as in the present case $I$. Also in our previous study \cite{EPJC71p1762}
of the $X(3872)$, with nine coupled channels, we reproduced the meson as a
confinement pole. What appears to happen in the present case $I$ \/is that
shifting the bare \ttpo\ state to somewhat lower energies is just enough to
deflect the confinement pole to the left and not the right
when approaching the continuum pole. Clearly, there will be an intermediate
situation for which the left/right deflection will hinge upon only marginal
changes in the parameters, but resulting in two completely different
trajectories. Therefore, identifying one pole as dynamical and the other
as linked to a confinement state is entirely arbitrary, the whole system being
dynamical because of unquenching. At the end of the day, the only thing that
really counts is where the poles end up for the final parameters. The
trajectories themselves are not observable and only serve as an illustration
how a coupled-channel model as the one employed here mimics the physical
situation. Suffice it to say that the lower pole, representing the $X(3872)$,
is quite stable with respect to variations in the parameters, owing to its
proximity to the only and most relevant OZI-allowed decay channel. The higher
pole, on the other hand, should not be taken at face value, since a more
realistic calculation should include other important decay channels,
such as $D^*D^*$, with threshold just above 4~GeV.

Concerning the other scenario with changed parameters, labeled $II$ \/in
Table~\ref{tab:newpar} and depicted in the lower graph of Fig.~\ref{newtraj},
we see that the trajectories do not change qualitatively when going from
the standard case to $II$. There is a displacement of the right-hand
branch, about 100~MeV to the right on average, in accordance with the same
shift of the bare \ttpo\ state. But the change in the lower, dynamical branch,
is much less significant, though the value of $g$ needed to produce a bound
state at 3871-68~MeV now increases to 1.572 (see Table~\ref{tab:newpar}).
We also notice from Fig.~\ref{newtraj} that the two pole-trajectory branches
hardly move towards one another, signaling less attraction between the
poles due to a larger initial separation.

Inspecting again Table~\ref{tab:newpar} as for the $c\bar{c}$ probability
in cases $I$ \/and $II$ \/compared to the standard situation, we observe an
increased value for case $I$ and a decreased one for $II$. This is logical,
since in case $I$ \/the bare \ttpo\ state lies closer to the $X(3872)$, whereas
in case $II$ \/it lies farther away. Nevertheless, the difference in $c\bar{c}$
probability between $I$ \/and $II$ \/is only about 3\%, i.e., less than the
variation with $a$ in the standard case. These comparisons lend further support
to the stability of our results.

\begin{figure}[t!]
\caption{Normalized two-component radial wave function $R(r)$, for cases $II$
\/and ''standard'', corresponding to parameters in Table~\ref{tab:newpar}. Bold
curves refer to case $II$, normal curve to $R_c$ for standard case. Note: $R_f$
is indistinguishable within graphical accuracy for the two cases.}
\mbox{ } \\[-30pt]
\hspace*{-20pt}
\resizebox{!}{400pt}{\includegraphics{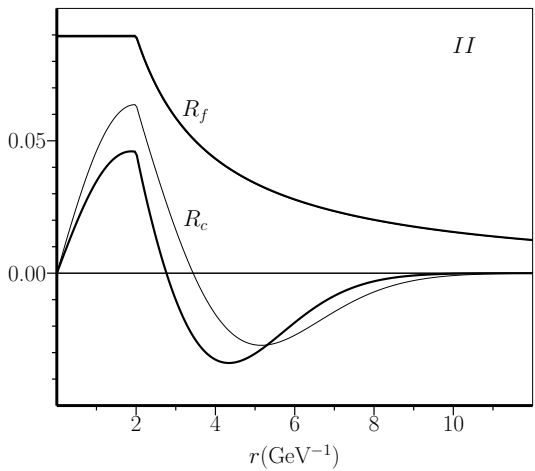}}\mbox{} \\[-200pt]
\label{nwf}
\end{figure}

Finally, in Fig.~\ref{nwf} we compare the wave function for case $II$ \/with the
standard one. We see there is no visible change in the $R_f$ component. 
As for $R_c$, the first maximum gets somewhat reduced, but the secondary,
negative bump even becomes a bit larger, owing to an inward shift of the node,
lying now at about 3~GeV$^{-1}$. Yet, also in case $II$ \/the $c\bar{c}$
component is still very significant, despite the large separation of more
than 200 MeV between the $X(3872)$ bound state and the bare \ttpo\ state.

From the latter and all previous results we may safely conclude that the
$c\bar{c}$ component of the $X(3872)$ wave function remains nonnegligible
in a variety of scenarios, being even of comparable size as the $D^0D^{*0}$
component in the inner region, save at very short distances.

\section{Summary and conclusions}  
\label{conc}
In the present paper, we have employed a simple and solvable
Schr\"odinger model to study the  wave function of the $X(3872)$ meson, by
treating it as a coupled $c\bar{c}$-$D^0D^{*0}$ system with $J^{PC}=1^{++}$
quantum numbers. Transitions between the two channels are described
with the \tpz\ mechanism, through string breaking at a sharp distance $a$.
The exact solutions to the equations allow us to easily study the
trajectories of $\mathcal{S}$-matrix poles as a function of the decay
coupling constant $g$. Thus, a dynamical pole is found, becoming a 
bound state just below the $D^0D^{*0}$ threshold, for different
string-breaking distances $a$, and an appropriate coupling $g$. On the other
hand, the pole arising from the bare \ttpo\ confinement state moves to 
higher energies and acquires a large imaginary part. However, the latter pole
may not be very relevant physically, because of neglected additional 
meson-meson channels which will become important in that energy
region.

As for the $X(3872)$ radial wave function, the $c\bar{c}$ component $R_c$
turns out to be of significant size as compared to the $D^0D^{*0}$ component
$R_f$, especially for intermediate $r$ values. Moreover, even for other trial
BEs, the global shape of $R_c$ and its relative magnitude vis-\`{a}-vis $R_f$
in the central region is remarkably stable. But the corresponding $c\bar{c}$
probability is relatively low, due to the very long tail of the $D^0D^{*0}$
wave function at small binding. These results are along the lines of the
analysis based on general arguments presented in Ref.~\cite{PPNP61p455}.
Quantitatively, for the average
\cite{PRD86p010001} $X(3872)$ binding of 0.16 MeV, a $c\bar{c}$ probability
of 7.5--11.2\% is found, for $a$ in the  range 0.4--0.6~fm, which is compatible
with other recent approaches \cite{PRD81p054023,1206.4877}.
The corresponding r.m.s.\ radius turns out to be quite stable at about 7.8~fm,
for the latter range of $a$ values, while the $S$-wave scattering length of
11.6~fm, for $a\approx0.4$~fm, is in agreement with expectations for a BE of
0.16 MeV. 

Finally, we have studied the nature of the $\mathcal{S}$-matrix pole giving 
rise to $X(3872)$, by varying some of the otherwise fixed parameters. Thus,
a drastic modification of pole trajectories is observed, for relatively
small parameter variations, making the $X(3872)$ pole transform from a
dynamical pole into one directly connected to the \ttpo\ bare confinement
state. However, the corresponding changes in the $c\bar{c}$ probability and
r.m.s.\ radius, as well as the coupling $g$ needed to reproduce $X(3872)$,
are quite modest.

In conclusion, we should revisit the claims about $X(3872)$ made in
Ref.~\cite{PRD76p094028}, quoted in the Introduction above, namely about the
inevitability of $X(3872)$ being a charm-meson molecule or virtual
state, independently of the mechanism generating the state. Now, it
is true that our analysis has confirmed some of the quantitative
predictions in Ref.~\cite{PRD76p094028}, viz.\ concerning the vanishing
probability of wave-function components other than $D^0D^{*0}$ as the
BE approaches zero, and the related behavior of the $D^0D^{*0}$ scattering
length. However, we have also shown that the $c\bar{c}$ component is certainly
not negligible and quite stable, in a variety of scenarios. Especially in
electromagnetic processes, the prominence of this component for relatively
small as well as intermediate $r$ values will no doubt result in a significant
contribution to the amplitudes. Moreover, as already mentioned above, the very
unquenching of a \ttpo\ $c\bar{c}$ state will not only introduce meson-meson
components into the wave function, but also a contribution of the \otpo\
$c\bar{c}$ state, which can change predictions of electromagnetic transition
rates very considerably \cite{PRD85p114002}. We intend to study such processes
for $X(3872)$ in future work, on the basis of a model as the one used in the
present paper, by employing the formalism developed and successfully applied
in Ref.~\cite{PRD44p2803}. However, in a detailed and predictive calculation
of electromagnetic $X(3872)$ decays, the inclusion of the charged
$D^\pm D^{*\mp}$ channel will be indispensable \cite{PRD81p014029}.

For all these reasons, we do not consider $X(3872)$
a charm-meson molecule, but rather a very strongly unitarized charmonium
state. As a matter of fact, we do not believe \em any \em \/non-exotic mesonic
resonance --- whatever its origin --- qualifies as a true meson-meson molecule,
simply because such a state will inexorably mix with the nearest $q\bar{q}$
states having the same quantum numbers. Indeed, we have demonstrated above
that, even with a bare $c\bar{c}$ state 200~MeV higher in mass, the resulting
$c\bar{c}$ component in the wave function is still appreciable. So let us
conclude this discussion by quoting and fully endorsing the following statement
from Ref.~\cite{PRD76p094028}:
\begin{quote} \em
``Any model of the $X(3872)$ that does not take into account its
strong coupling to charm meson scattering states should not be taken seriously.''
\em
\end{quote}

\begin{acknowledgement}
One of us (G.R.) is indebted to R.~M.~Woloshyn for the invitation to a most
stimulating miniworkshop at TRIUMF last year, where the topic leading to
the present paper was debated. Thanks are also due to E.~Braaten and
K.~K.~Seth for very useful discussions on $X(3872)$.
This work was partially supported by
the \emph{Funda\c{c}\~{a}o para a Ci\^{e}ncia e a Tecnologia}
\/of the \emph{Mi\-nist\'{e}rio da Educa\c{c}\~{a}o e Ci\^{e}ncia}
\/of Portugal, under contract no.\ CERN/FP/123576/2011.
\end{acknowledgement}

\appendix
\section{Solving the coupled-channel Schr\"odinger equation}
\label{appA}
We twice integrate the Schr\"odinger equation~\eqref{schr} in order to get
two sets of boundary conditions, viz.\ Eqs.~(\ref{bc1}) and (\ref{bc2}):
\be
\label{bc1}
\begin{split}
&u_c'(r\upr a)-u_c'(r\dar a)+\frac{g}{a}u_f(a)=0 \; ,\\
&u_f'(r\upr a)-u_f'(r\dar a)+\frac{g\mu_f}{a\mu_c}u_c(a)=0 \; ;
\end{split}
\ee
\be
\begin{split}
\label{bc2}
&u_c(r\upr a)=u_c(r\dar a) \; , \\
&u_f(r\upr a)=u_f(r\dar a) \; .
\end{split}
\ee\mbox{} \\[2mm]
A general solution to this problem is the two-component wave function given by
Eqs.~(\ref{wfc}) and \eqref{fwf}, for the confined and meson-meson channel, 
respectively:
\be
\label{wfc}
u_c(r)=
\left\lbrace\barr{lc}
A_cF_c(r) & r<a \; , \\[5pt]
B_cG_c(r) & r>a \; ;
\earr\right.
\ee
\be
\label{fwf}
u_f(r)=\left\lbrace\barr{lc}
A_f J_{l_f}(kr)\; , & r<a \; , \\[5pt]
B_f\Big\lbrack J_{l_f}(kr)k^{2l_f+1}\\
\hspace*{20pt}\cot\delta_{l_f}(E)-N_{l_f}(kr)\Big\rbrack \; , &r>a\;.
\earr\right.
\ee
In Eq.~(\ref{wfc}), the function $F_c(r)$ vanishes at the origin, whereas
$G_c(r)$ falls off exponentially for $r\to\infty$, their explicit expressions
being
\begin{eqnarray}
\label{ffc}
\hspace*{-20pt} F(r)=&\displaystyle\:\frac{1}{\Gamma(l+3/2)}\,z^{(l+1)/2}\,
e^{-z/2}\,\Phi(-\nu,l+3/2,z) \; , & \\
\displaystyle
\label{fgc}
\hspace*{-20pt} G(r)=&\displaystyle-\frac{1}{2}\Gamma(-\nu)\,z^{l/2}\,e^{-z/2}
\,\Psi(-\nu,l+3/2,z) \; ,  &
\end{eqnarray}
where $\Phi$ and $\Psi$ are the confluent hypergeometric functions of first and
second kind (see Appendix \ref{appB}), respectively, $\Gamma(-\nu)$ is the
complex gamma function, $\nu$ is given by Eq.~(\ref{eve}), and
$z=\mu\omega r^2$. Note that only in the case of integer $\nu$, i.e., for 
$g=0$, do $\Phi$ and $\Psi$ reduce to the usual Laguerre polynomials for the
three-dimensional HO potential.
Furthermore, the functions $J$ and $N$ in Eq.~(\ref{fwf}) are simple
redefinitions of the standard spherical Bessel and Neumann functions, i.e.,
$J_l(kr)=k^{-l}rj_l(kr)$ and $N_l(kr)=k^{l+1}rn_l(kr)$.
From the boundary conditions (\ref{bc1}) and (\ref{bc2}), as well as the
wave-function expressions (\ref{wfc}) and (\ref{fwf}), we get, with the
definition $\kappa=ka$,
\begin{eqnarray}
\label{bc1b}
\displaystyle
G_c'(r)F_c(a)-F_c'(a)G_c(a)&=&
\frac{g}{a}J_{l_f}(\kappa)F_c(a)\frac{A_f}{B_c} \; , \nonumber \\
\mbox{ } \\
\displaystyle
J_{l_f}'(\kappa)N_{l_f}(\kappa)-J_{l_f}(\kappa)N_{l_f}'(\kappa)&=&
\frac{g}{a}\frac{\mu_f}{\mu_c}J_{l_f}(\kappa)F_c(a)\frac{A_c}{B_f} \; .
\nonumber
\end{eqnarray}
Using now the Wronskian relations
\begin{eqnarray}
\label{bc2b}
\hspace*{-10pt} W(F_c(a),G_c(a)) \equiv F_c(a)G_c'(a)-F_c'(a)G_c(a)=1\;,
 \nonumber \\ \mbox{ } \hspace{-20pt} \\
W(N_{l_f}(\kappa),J_{l_f}(\kappa)) \equiv N_{l_f}(\kappa)J_{l_f}'(\kappa) -
N_{l_f}'(\kappa)J_{l_f}(\kappa)=-1 \; , \hspace{-15pt} \nonumber
\end{eqnarray}
and continuity of the wave function at
$r\!=\!a$ (cf.\ Eq.~(\ref{bc2})), we can solve for three of the four unknowns
$A_c$, $B_c$, $A_f$, and $B_f$. Note that Eqs.~(\ref{bc1}) and (\ref{bc2})
are not entirely linearly independent, so that solving all four constants
is not possible. This is logical, as the overall wave-function normalization
does not follow from the Schr\"{o}dinger equation. Expressing all in terms of
$A_c$ then yields
\be
\label{amp}
\begin{array}{cc}
A_c \; , &
\displaystyle A_f=-\left[\frac{g}{a}J_{l_f}(\kappa)G_c(a)\right]^{-1}\!A_c\;,
\\[10pt] \displaystyle
B_c=\frac{F_c(a)}{G_c(a)}\,A_c \; , &
\displaystyle B_f=\frac{g}{a}\frac{\mu_f}{\mu_c}J_{l_f}(\kappa)F_c(a)\,A_c\;.
\end{array}
\ee
Note that, in order to obtain the $D^0D^{*0}$ wave function in the outer
region, we must substitute $\cot\delta_{l_f}(E)=i$ in
Eq.~(\ref{fwf}) (also see below). Finally, the normalization constant
$\mathcal{N}$ of the total wave function is determined by computing
\be
\label{n}
\int_0^\infty \!dr\ |u(r)|^2=
\int_0^\infty \!dr\ \left(u_c^2(r)+u_f^2(r)\right)=\mathcal{N}^2 \; .
\ee
Then, we can also calculate the root-mean-square radius
$\bar{r}=\sqrt{\langle r^2\rangle}$ of the two-component system by
\be
\label{rms}
\langle r^2\rangle = \frac{1}{\mathcal{N}^2}\int_0^\infty dr\,r^2
\left(u_c^2(r)+u_f^2(r)\right) \; .
\ee

As for the $\mathcal{S}$-matrix poles corresponding to resonances, bound
states, or virtual bound states, $\cot\delta_{l_f}(E)$ can be
solved from continuity of $u_f(r)$ at $r\!=\!a$ in Eq.~(\ref{fwf}),
resulting in the expression
\be
\label{cotan}
\cot\delta_{l_f}(E)=
-\left[g^2\frac{\mu_f}{\mu_c}kj_{l_f}^2(\kappa) F_c(a)G_c(a)\right]^{-1}+
\frac{n_{l_f}(\kappa)}{j_{l_f}(\kappa)} \; ,
\ee
with the $1\times1$ $\mathcal{S}$-matrix simply given by
\be
\label{scotan}
S_{l_f}(E)=\frac{\cot\delta_{l_f}(E)+i}
{\cot\delta_{l_f}(E)-i} \; .
\ee
Real or complex poles can then be searched for numerically, by using
Newton's method to find the energies for which
$\cot\delta_{l_f}(E)=i$, on the appropriate Riemann sheet.

\section{Special functions, numerical methods, and kinematics}
\label{appB}
The confluent hypergeometric functions $\Phi$ and $\Psi$ introduced in
Appendix~\ref{appA} are defined in Ref.~\cite{B53}, Eqs.~(6.1.1) and
(6.5.7), respectively. Thus, the function $\Phi$ is easily programmed as 
a rapidly converging power series, while the definition (6.5.7) of $\Psi$
in terms of $\Phi$ and the gamma function $\Gamma$ then also allows
straightforward computation, by employing Gauss's multiplication formula
for $\Gamma(-\nu)$ (see Ref.~\cite{AS70}, Eq.~(6.1.20)) so as to map the
argument $-\nu$ to lying well inside the unit circle in the complex plane,
whereafter a very fast converging power-series expansion of
$1/\Gamma(-\nu)$ (see Ref.~\cite{AS70}, Eq.~(6.1.34)) can be applied.

The integrals for wave-function normalization and computation of r.m.s.\
radii are carried out by simple Gauss integration, choosing increasing
numbers of points on a finite interval for the $c\bar{c}$ channel, and an
infinite one for $D^0D^{*0}$. Note that, in the former case, the wave
function falls off fast enough to allow convergence for a finite cutoff,
whereas in the latter a suitable logarithmic mapping is used. In both
cases though, because of the wave-function cusp at $r\!=\!a$ and in order
to avoid numerical instabilities, the domain of integration is split into
two pieces, with up to 16 Gauss points in the inner region and 64 in the
outer one, thus resulting in a very high precision of the results.

Although the $X(3872)$ bound state can reasonably be considered a
nonrelativistic system, we still use relativistic kinematics in the
$D^0D^{*0}$ channel, since parts of the resonance-pole trajectories
involve relatively large (complex) momenta. For consistency, the same
is done for all energies. The manifest unitarity of the $\mathcal{S}$
matrix is not affected by this choice. Thus, the relative $D^0D^{*0}$
momentum reads
\be
\label{mom}
k(E)=\frac{E}{2}\left\lbrace\left[1-\left(\frac{T}{E}\right)^2\right]
\left[1-\left(\frac{P}{E}\right)^2\right]\right\rbrace^{\frac{1}{2}} \; ,
\ee
where $T$ and $P$ are the threshold ($m_{D^{*0}}+m_{D^0}$) and
pseudothreshold ($m_{D^{*0}}-m_{D^0}$) energies, respectively.
The corresponding relativistic reduced mass is defined as
\be
\label{redm}
\mu_f(E)\equiv\frac{1}{2}\frac{dk^2}{dE}=
\frac{E}{4}\left[1-\left(\frac{TP}{E^2}\right)^2\right] \; .
\ee
Note that in the $c\bar{c}$ channel the reduced mass is defined in the
usual way, i.e., $\mu_c=m_c/2$, owing to the inherently nonrelativistic
nature of the HO potential and the ensuing wave function.

\end{document}